\documentclass[aps,pre,twocolumn,groupedaddress,showpacs]{revtex4-1}

\usepackage{natbib}
\usepackage{graphics}
\usepackage{amsfonts}
\usepackage{amssymb}
\usepackage{amsmath}

\begin{document}

\title{Revisiting the phase diagram of hard ellipsoids}

\author{Gerardo Odriozola}
\email{email: godriozo@imp.mx} 
\affiliation{Programa de Ingenier\'{\i}a
Molecular, Instituto Mexicano del Petr\'{o}leo, Eje Central
L\'{a}zaro C\'ardenas 152, 07730, M\'{e}xico, Distrito Federal,
M\'{e}xico.}

\date{\today}

\begin{abstract}
In this work the well-known Frenkel-Mulder phase diagram of hard ellipsoids of revolution [Mol. Phys. 55, 1171 (1985)] is revisited by means of replica exchange Monte Carlo simulations. The method provides good sampling of dense systems and so, solid phases can be accessed without the need of imposing a given structure. At high densities, we found plastic solids and fcc-like crystals for semi-spherical ellipsoids (prolates and oblates), and SM2 structures [Phys. Rev. E 75, 020402 (2007)] for $x\!:\!1$-prolates and $1\!:\!x$-oblates with x$\geq$3. The revised fluid-crystal and isotropic-nematic transitions reasonably agree with those presented in the Frenkel-Mulder diagram. An interesting result is that, for small system sizes (100 particles), we obtained 2:1 and 1.5:1-prolate equations of state without transitions, while some order is developed at large densities. Furthermore, the symmetric oblate cases are also reluctant to form ordered phases. 
\end{abstract}


\maketitle

\section{Introduction}

As early as in 1984\footnote{It was early in terms of computing power. Frenkel et al.~works~\cite{Frenkel84,Frenkel85} were done with about a hundred times less computing power than that of a current \textquotedblleft smart\textquotedblright cell-phone}, Frenkel et al.~provided a tentative picture of the phase diagram of hard ellipsoids of revolution~\cite{Frenkel84}. 
This was the base of the well-known Frenkel-Mulder complete phase diagram (FMD) published the following year~\cite{Frenkel85}. It basically consists of four phases: solid (assumed to be a stretched-fcc), plastic solid (a solid constituted of orientationally disordered particles), nematic fluid, and isotropic fluid. These phases appear at different regions of the density - aspect ratio plane. Naturally, fluid and solid phases locate at low and high densities, respectively, whereas long range orientational correlations are favored for increasing asymmetry. For increasing density and sufficiently asymmetric particles, the orientational order develops before the positional order. Conversely, for semi-spherical particles, the positional order appears at lower densities than the orientational order. This behavior is observed for both, oblates and prolates. Consequently, the FMD is highly symmetric around the spherical case.        

The FMD has become the starting point for many studies on uniaxial hard ellipsoids~\cite{Allen95,Camp96,Vega97,Tjipto90,DeMichele07}. In addition, this phase diagram has been practically unchanged until quite recently. One of these recently occurred changes is that the line of maximally achievable density has moved upwards~\cite{Donev04a}. This is due to the fact that it was found a family of crystal structures (SM2~\cite{Donev04a, Radu09}) which produces maximum packing densities well above the stretched-fcc crystal (it has the same maximum packing density as the fcc~\cite{Frenkel85,TorquatoRev}). A second and related change is that for relatively large asymmetry  and densities close to the fluid-crystal transition, this family of crystal structures has a lower free energy than the stretched-fcc imposed by Frenkel et al.~in their calculations~\cite{Radu09} (this does not occur for semi-spherical particles). Consequently, different solid crystal regions naturally appear in the phase diagram~\cite{Radu09,Pfleiderer07}. Furthermore, these free energy differences between the solid structures raise questions on the accuracy of the locations of the fluid-crystal transitions in the FMD. 

In this paper we revisit the phase diagram of hard uniaxial ellipsoids. The aim is twofold. First, we would like to further investigate whether or not the SM2 structure is effectively the one which corresponds to equilibrium. Second, we would like to relocate the transitions without imposing any a priori crystal structure. For these purposes, we implemented the replica exchange Monte Carlo (REMC) method since it has the advantage of naturally producing ordered structures from random configurations~\cite{OdriozolaHE,GuevaraHE}. The price to be paid is that only small systems can be explored\footnote{Five quad-cores were used for this study.}. We found that the algorithm yields plastic solids and fcc-like crystals for semi-spherical ellipsoids (prolates and oblates), and SM2 structures for $x\!:\!1$-prolates and $1\!:\!x$-oblates with x$\geq$3. The locations of the transitions, however, reasonable agree with those presented in the FMD. Another interesting result is that, for the studied system sizes (100 particles), we obtained 2:1 and 1.5:1-prolate equations of state without signs of transitions, while some order is developed at large densities. 1:2 and 1:1.5-oblates were also found to hinder the formation of ordered phases.  

The paper is organized in four sections. The first one is this brief introduction. In the second the employed models and methods are described. In the next one, the equations of state for different prolate and oblate systems are presented. At the end of this same section the phase diagram summarizes the results. Conclusions are drawn in a final section.    

\section{Models and methods}

\subsection{Rickayzen-Berne-Pechukas hard ellipsoids}

Rickayzen introduced a modification of the well established Berne and Pechukas~\cite{Berne72} (BP) (also called hard Gaussian overlap~\cite{Varga02}) model. Its aim is to give a better analytical approach, by fixing the T-shape BP mismatch, to the exact solution of the distance of closest approach of a pair ($i$ and $j$) of uniaxial ellipsoids. The Rickayzen-Berne-Pechukas (RBP) expression reads~\cite{Rickayzen98}
\begin{equation}\label{RBP1}
\sigma_{RBP} = \sigma_{\bot} \bigg ( 1 - \frac{1}{2} \chi \big [ A^{+} + A^{-} \big ] + \big ( 1 - \chi ) \chi' \big [ A^{+} A^{-} \big ]^{\gamma} \bigg )^{-1/2}
\end{equation}
being
\begin{equation}\label{RBP2}
A^{\pm} = \frac{ ( \hat{\mathbf{r}} \cdot \hat{\mathbf{u}}_{i} \pm \hat{\mathbf{r}} \cdot \hat{\mathbf{u}}_{j} )^{2} }{ 1 \pm \chi \hat{\mathbf{u}}_{i} \cdot \hat{\mathbf{u}}_{j} },
\end{equation}
\begin{equation}\label{RP2}
\chi = \frac{ \sigma_{\|}^{2} - \sigma_{\bot}^{2} }{ \sigma_{\|}^{2} + \sigma_{\bot}^{2} },
\end{equation}
and 
\begin{equation}\label{RBP3}
\chi' = \bigg ( \frac{ \sigma_{\|} - \sigma_{\bot} }{ \sigma_{\|} + \sigma_{\bot} } \bigg )^{2}.
\end{equation}
Here, $\sigma_{\|}$ and $\sigma_{\bot}$ are the parallel and perpendicular diameters with respect to the main ellipsoid axis, respectively. We are following the (redundant but clear) notation $1\!:\!x$-oblates ($x=\sigma_{\bot}$) and $x\!:\!1$-prolates ($x=\sigma_{\|}$) to denote the oblates and prolates aspect ratios. $\hat{\mathbf{u}}_{i}$ and $\hat{\mathbf{u}}_{j}$ are the unit vectors along the main axis of each particle, and $\hat{\mathbf{r}}$ is the unit vector along the line joining the geometric centers. Furthermore, $\gamma$ is introduced~\cite{GuevaraHE} as a free adjusting parameter to approach even further the exact Perram and Wertheim (PW) numerical solution~\cite{Perram84,Perram85} ($\gamma=1$ also produces good results~\cite{GuevaraHE}). This expression, with the $\gamma$ values given in reference~\cite{GuevaraHE}, is the model we are employing to approach the ellipsoidal entities in our simulations. This implementation is close to three times faster than using the exact PW iterative solution~\cite{GuevaraHE}. As was shown for $x\leq5$, the average difference between RBP and the exact numerical solution for a collection of $10^8$ random configurations (varying $\hat{\mathbf{r}}$, $\hat{\mathbf{u}}_{i}$, and $\hat{\mathbf{u}}_{j}$) is always below $0.8\%$. The equations of state corresponding to RBP and PW were also compared showing no practical differences for 1:5-oblates~\cite{GuevaraHE}.

\subsection{Replica Exchange Monte Carlo}

As mentioned in the introduction, the REMC method has the capability of producing high density ordered structures from low density random configurations by using a reasonable amount of computing power. Moreover, it can sample the high density configuration space by producing many different crystal structures (weighted by their corresponding factors as in a standard MC simulation). In other words, this technique is well proven to assist the systems to reach equilibrium at difficult (high density / low temperature) conditions~\cite{Marinari92,Lyubartsev92,hukushima96}. Thus, there is no need of imposing a given crystal structure (with the hope it is representative of equilibrium) to produce the equation of state of a solid phase, on the one hand, and on the other, the produced ordered phases can be analyzed. 

The method is based on the definition of an extended ensemble whose partition function is given by $Q_{extended}=\prod_{i=1}^{n_r}Q_{i}$, being $n_r$ the number of ensembles and $Q_i$ the partition function of ensemble $i$. This extended ensemble is sampled by setting $n_r$ replicas, each replica placed at each ensemble. The sampling of each ensemble can be performed as in a standard MC, while non standard trials and biased sampling strategies can also be implemented to further increase performance. The existence of the extended ensemble justifies the introduction of swap trial moves between any two replicas, whenever the detail balance condition is satisfied. Since we are dealing with hard particles, it is convenient to make use of isobaric-isothermal ensembles and perform the ensemble expansion in pressure~\cite{Odriozola09} (this same procedure is suitable for any kind of purely repulsive interactions). Hence, the partition function of the extended ensemble is given by~\cite{Okabe01,Odriozola09} 
\begin{equation}
Q_{\rm extended}=\prod_{i=1}^{n_r} Q_{N T P_i},
\end{equation} 
where $Q_{NTP_i}$ is the partition function of the isobaric-isothermal ensemble of the system at pressure $P_i$, temperature $T$, and with $N$ particles. 

We implemented a standard sampling of the $NTP_i$ ensembles. This involves independent trial displacements, rotations of single ellipsoids, and volume changes. To increase the degrees of freedom of our small systems ($N=100$), we implemented non-orthogonal parallelepiped cells. Thus, sampling also includes trial changes of the angles of the lattice vectors. This is done while rescaling the cell sides and particles positions to preserve volume and keep a simple acceptation rule. We observed that at high densities and for certain conditions the introduction of non-orthogonal cells improves the results (the crystal branch slightly shifts to larger densities). Swap moves are performed by selecting at random a given adjacent pair of replicas and using the following acceptance probability~\cite{Odriozola09}
\begin{equation}
\label{accP} 
P_{\rm acc}\!=\! \min(1,\exp[\beta(P_i-P_j)(V_i-V_j)]), 
\end{equation} 
where $\beta=1/(k_BT)$ is the reciprocal temperature and $V_i-V_j$ is the volume difference between replicas $i$ and $j$. Adjacent pressures should be close enough to provide reasonable swap acceptance rates between neighboring ensembles. In order to take good advantage of the method, the ensemble at the smaller pressure must also ensure large jumps in configuration space, so that the higher pressure ensembles can be efficiently sampled.

Simulations are started by randomly placing the ellipsoids with random orientations (avoiding overlaps), so that the initial volume fraction is $\varphi= v_{e} \rho = 0.2$, where $\rho$ is the number density and $v_{e}=4 \pi \sigma_{\|} \sigma_{\bot}^2/3$ is the ellipsoid volume. We first perform about $2 \times 10^{13}$ trial moves at the desired state points, during which we observe the replicas reaching a stationary state (equilibrating procedure). We then perform $2 \times 10^{13}$ additional sampling trials. The maximum particle displacements, maximum rotational displacements, maximum volume changes, and maximum changes of the angles of the lattice vectors are adjusted for each pressure to yield acceptance rates close to 0.3. Since an optimal allocation of the replicas should lead to a constant swap acceptance rate for all pairs of adjacent ensembles~\cite{Rathore05}, we implemented a simple algorithm to smoothly adjust the intermediate pressures while keeping the maximum and minimum pressures fixed. To start the simulations, we used a geometric progression of the pressure with the replica index. These adjusting procedures are performed only during the equilibrating stage (the first $2 \times 10^{13}$ trial moves). Verlet neighbor lists~\cite{Donev05a,Donev05b} are used to improve performance. 
We set $N=100$ ellipsoids and $n_r=64$ (to cover a wide range of densities while keeping large swap acceptance rates). More details are given in previous work~\cite{GuevaraHE}.

\section{Results}

\begin{figure}
\resizebox{0.45\textwidth}{!}{\includegraphics{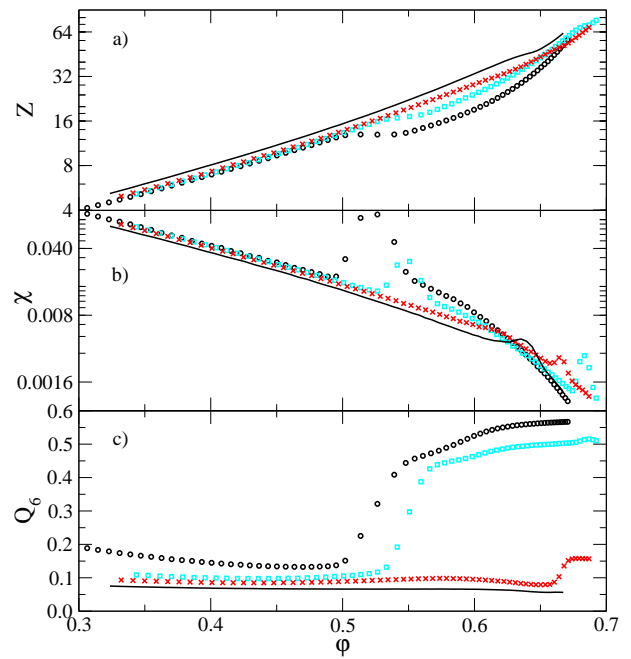}}
\caption{\label{Fig1} a) Equations of state for oblates, $Z(\varphi)$. b) Isothermal compressibility, $\chi(\varphi)$. c) Order parameter, $Q_6(\varphi)$. For all panels, black (dark) circles, cyan (light gray) squares, red (gray) crosses, and the solid line correspond to $1\!:\!x$-oblates with $x=1.1$, 1.3, 1.5, and 2.0, respectively. }
\end{figure}

The dimensionless pressure, $Z=\beta P/\rho$, as a function of the volume density, $\varphi$, (equation of state) for $1\!:\!x$-oblates with $x=1.1$, 1.3, 1.5, and 2.0 is shown in panel a) of figure~\ref{Fig1}. The corresponding isothermal compressibility, as obtained from density fluctuations, $\chi=N(<\rho^2>-<\rho>^2)/<\rho>^2$, is given in panel b) of the same figure. Finally, the order parameter $Q_6$ is shown in panel c). This parameter is given by $Q_6=\left(\frac{4\pi}{13}\sum_{m=-6}^{m=6}|<\!Y_{6m}(\theta,\phi)\!>|^2\right)^{1/2}$, being $<\!Y_{6m}(\theta, \phi)\!>$ the ensemble average over all bonds of the spherical harmonics of the polar angles $\theta$ and $\phi$~\cite{Steinhardt96,Rintoul96b,TorquatoRev}. The bonds are considered as the vectors joining the geometrical centers of two neighboring particles. $Q_6(\varphi)$ approaches zero for a completely random system of a large number of points, and increases when configurations present positional order. 

1:1.1-oblates are semi-spherical particles. Consequently, the system is expected to behave similarly to hard-spheres (HS). That is, we should obtain an isotropic fluid phase at low densities, a crystal phase at large densities, and a transition located close to the HS transition. Since asymmetry is very week, the crystal phase should not show long range orientational correlations, unless very high pressures are applied (we do not expect to achieve this orientational ordered phase in this case). This is exactly what black circles of figure~\ref{Fig1} show. $Z(\varphi)$ shows two branches divided by a $\varphi$ jump, pointing out a first order transition. This region also produces large values of $\chi(\varphi)$, which should diverge at the thermodynamic limit, and a strong increase of $Q_6(\varphi)$ to reach values very close to the fcc limit~\cite{TorquatoRev} (0.5745~\cite{Rintoul96b}). We also observed $p(r)=<1/2(3(\hat{\mathbf{u}}_{i} \cdot \hat{\mathbf{u}}_{j})^2-1)>$ with values close to zero (meaning no orientational correlations) at long enough distances, $r$. These two last findings, together with a snapshot inspection (see the plastic crystal inset of figure~\ref{Fig5}), point out that the solid branch corresponds to a plastic solid. The transition is placed slightly to the right as compared to the HS transition obtained with the same algorithm and system size (see table~\ref{Table1}). These transitions (fluid-solid) are expected to shift to the right, get wider, and be produced at slightly higher pressures for the thermodynamic limit~\cite{Odriozola09,GuevaraHE}.

When oblates asymmetry is increased to 1:1.3, cyan squares of figure~\ref{Fig1}, most features observed for the 1:1.1 case remain. That is, there is an isotropic fluid region, a plastic crystal branch, and the corresponding transition. The quantitative differences are that the transition shifts to the right, more pressure is needed to produce the plastic solid phase, and that the $\varphi$ jump turns smaller. Additionally, at very high pressures, a second and small $\varphi$ jump is captured. This jump occurs along with the development of a local maximum of $\chi(\varphi)$ (panel b) of figure~\ref{Fig1}) and a practically constant and large $Q_6(\varphi)$. An inspection of function $p(r)$ (not shown) and snapshots (see the prolate, fcc-like crystal of figure~\ref{Fig6}) indicates that long range (relative to the sides of the small cells we are employing) orientational correlations are developed for $\varphi>0.682$. Thus, the algorithm is probably capturing the plastic solid- fcc transition.  

\begin{figure}
\resizebox{0.45\textwidth}{!}{\includegraphics{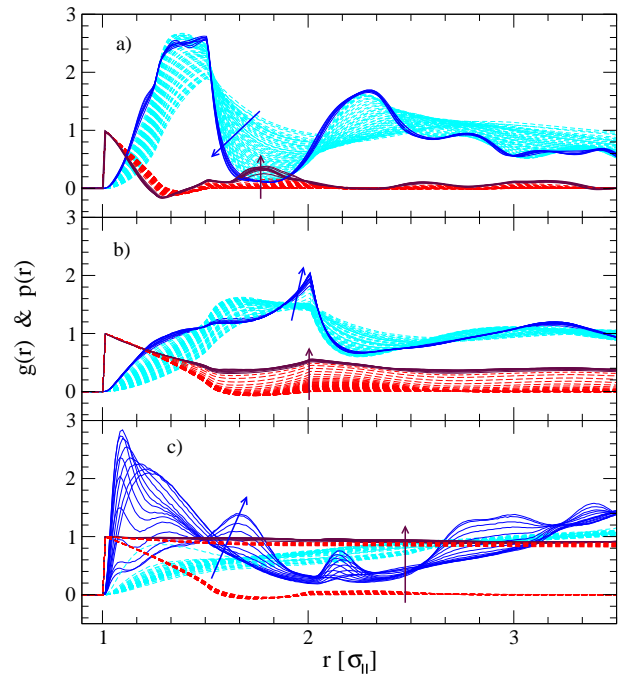}}
\caption{\label{Fig2} Radial distribution functions ($g(r)$), cyan (light gray) dashed and blue (dark) solid lines, and their corresponding radial orientational order parameters ($p(r)$), red (gray) dashed and dark red solid lines, for all pressures (increasing as indicated by the arrows). Panels a), b), and c) correspond to $1\!:\!x$-oblates with $x=1.5$, 2.0, and 4.0, respectively. Cyan (light gray) dashed and red (gray) curves correspond to low density states. Blue (dark) solid and dark red correspond to solid states in panels a) and c) and to nematic states in panel b).}
\end{figure}

The 1:1.5-oblate system is reluctant to crystallize, but it does. This occurs at large densities and high pressures and the obtained ordered phases do not show a clear structure. In fact, $Q_6(\varphi)$ points out the existence of some, but not large, positional order. Anyway, a tiny $\varphi$ jump appears (unnoticeable from panel a) of figure~\ref{Fig1}) along with a $\chi(\varphi)$ maximum and a $Q_6(\varphi)$ sharp increase. All these features are seen from the red crosses of figure~\ref{Fig1}. At this transition $p(r)$ does not show long range orientational order (panel a) of figure~\ref{Fig2}). Before the transition, at $\varphi \approx 0.62$, a change in the slope of $\chi(\varphi)$ should be noted. This change occurs where the isotropic fluid shows extremely low translational and rotational diffusion coefficients~\cite{DeMichele07,Pfleiderer08a,Pfleiderer08b}, i.~e. dynamics turns glassy, and it may suggest a high order transition~\cite{Hermes10,PerezAngel11,Odriozola11}.

The reluctance to form a crystal phase persists for 1:2.0-oblates (black lines of figure~\ref{Fig1}). Here, $Q_6(\varphi)$ does not point out any sign of positional order. Nonetheless, both, $Z(\varphi)$ and $\chi(\varphi)$ indicate a transition at $\varphi \approx 0.634$. An inspection of function $p(r)$ shows the development of long range orientational order at this density (see panel b) of figure~\ref{Fig2}). Thus, according to these data, we must label this transition as isotropic-nematic (fluid-fluid). This point makes continuous the isotropic-nematic transition line (blue dashed line of figure~\ref{Fig5}). We were, however, also expecting a fluid-solid transition. A fluid-solid transition would not only have sense by avoiding an interruption of the fluid-solid coexistence region (see figure~\ref{Fig5}), but it would also agree with previously reported data~\cite{Frenkel85}.  

Before moving to more asymmetric oblate cases, we should present the radial distribution functions, $g(r)$, and the radial orientational order parameters, $p(r)$, for the 1:1.5 and 1:2-oblate cases. These are shown in panels a) and b) of figure~\ref{Fig2}, respectively. In panel a) the cyan dashed and blue solid lines correspond to the $g(r)$ functions before and after the fluid-solid transition, respectively. The inserted blue arrow shows the direction of increasing pressure. Blue solid lines point out more structured systems than cyan dashed lines (peaks and valleys are more pronounced and better defined) although changes are not very remarkable. Indeed, differences between the highest pressure cyan dashed line and the lowest pressure blue solid line are not large, as they usually are for state points before and after a fluid-solid transition. This is in agreement with the relatively low $Q_6(\varphi)$ values found for the solid. On the other hand, and as was already mentioned, $p(r)$ functions show no signs of development of long range orientational order. At the transition, however, a small $p(r)$ peak grows at the $g(r)$ valley. The 1:2-oblate case of panel b) shows cyan dashed and blue solid lines ($g(r)$ functions) corresponding to states before and after the isotropic-nematic transition. Aside the peak growing at $r=\sigma_{\bot}$ (as indicated by the blue arrow of panel b)), curves are smooth, signaling a structureless fluid phase. The peak at $r=\sigma_{\bot}$ is linked to side-to-side configurations appearing as a consequence of a layering process~\cite{OdriozolaHE,GuevaraHE}. This layering process is also pointed out by the $p(r)$ functions. Here, red dashed and dark red solid lines correspond to states before and after the isotropic-nematic transition. The long range orientational order is clearly shown by the dark red solid lines. Finally, figure~\ref{Fig2} c) shows the 1:4-oblate case. In this case, the development of a crystalline structure is evident at the fluid-solid transition (dashed cyan and red lines correspond to the $g(r)$ and $p(r)$ functions before the fluid-solid transition, whereas blue solid and dark red lines represent the states after this transition). Additionally, there are red dashed lines showing long range orientational correlations at lower pressures than for the fluid-solid transition, pointing out the presence of a fluid-nematic phase.  

\begin{figure}
\resizebox{0.46\textwidth}{!}{\includegraphics{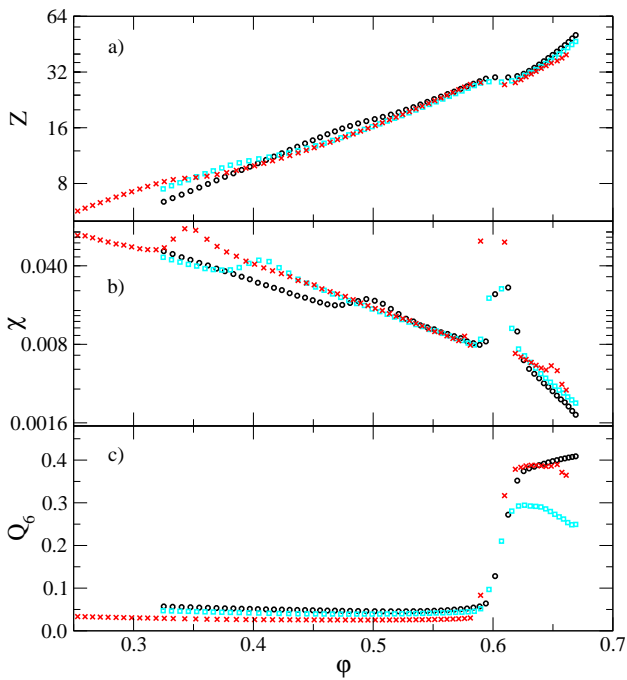}}
\caption{\label{Fig3} Equations of state for oblates, $Z(\varphi)$. b) Isothermal compressibility, $\chi(\varphi)$. c) Order parameter, $Q_6(\varphi)$. For all panels, black (dark) circles, cyan (light gray) squares, and red (gray) crosses correspond to $1\!:\!x$-oblates with $x=3.0$, 4.0, and 5.0 (taken from reference~\cite{GuevaraHE}), respectively.}
\end{figure}

Figure~\ref{Fig3} shows $Z(\varphi)$, $\chi(\varphi)$, and $Q_6(\varphi)$ for $1\!:\!x$-oblates with $x=3.0$, 4.0, and 5.0. These cases are clear, i. e., there is high consistency among all defined measurables and with previous works. All point out a low pressure isotropic branch, followed by an intermediate pressure nematic branch, and end with a high pressure solid region. Thus, these cases have a low pressure isotropic-nematic transition followed by a high pressure nematic-crystal transition. Their locations depend on the aspect ratio (see table~\ref{Table1}). In brief, increasing asymmetry leads to a shift of both transitions to lower pressures and densities (ordered phases are favored). This shifting is much more pronounced for the fluid-fluid transition though (see panel b) of figure~\ref{Fig3}). A curious fact is that $Z(\varphi)$ and $\chi(\varphi)$ look independent of the aspect ratio for the nematic phase. Nonetheless, they do not for the solid branch. This contrast with the fact that these branches are expected to diverge at $\varphi=0.7707...$~\cite{Donev04a}. This is probably a consequence of the presence of size effects and/or differences between the model and the exact ellipsoidal shape~\cite{GuevaraHE}.     

\begin{figure}
\resizebox{0.45\textwidth}{!}{\includegraphics{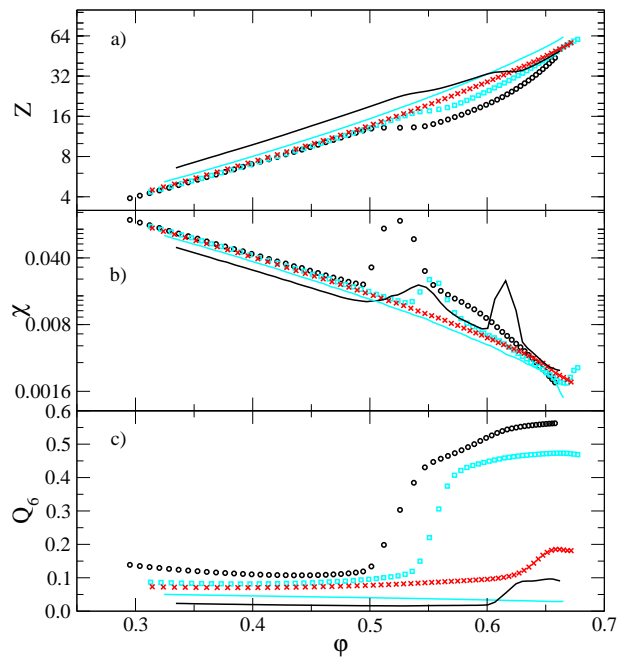}}
\caption{\label{Fig4} Equations of state for prolates, $Z(\varphi)$. b) Isothermal compressibility, $\chi(\varphi)$. c) Order parameter, $Q_6(\varphi)$. For all panels, black (dark) circles, cyan (light gray) squares, red (gray) crosses, the black (dark) solid line, and the cyan (light gray) solid line correspond to $x\!:\!1$-prolates with $x=1.1$, 1.3, 1.5, 2.0, and 3.0, respectively.}
\end{figure}

The equations of state of $x\!:\!1$-prolates, with $x=1.1$, 1.3, 1.5, 2.0, and 3.0, are shown in figure~\ref{Fig4}. Performing the same  analysis than for oblates, that is, in base of functions $Z(\varphi)$, $\chi(\varphi)$, $Q_6(\varphi)$, $g(r)$, $p(r)$, and a snapshot analysis, we reach to the following: For $x\leq1.3$ there is an isotropic-plastic solid phase transition. The expected~\cite{Frenkel85} (but hard to capture) plastic solid- fcc transition is detected only for the $x=1.3$ case (a snapshot of the prolate, fcc-like solid is shown in figure~\ref{Fig6}). For $x=3.0$ we found the low density isotropic-nematic and high density fluid-solid transitions (only the low density region was studied for cases with $x\geq4.0$, since needle-like particles turn larger than half the cell side at high pressures for $N=100$). Thus, as was already pointed out~\cite{Frenkel85}, the phase diagram is highly symmetric (prolates and oblates behave very similarly). Also keeping symmetry, not very conclusive results are yield in the region $1.3<x<3.0$. For $1.5:1$ and $2.0:1$-prolates, there appear no signs of transitions in $Z(\varphi)$ or in $\chi(\varphi)$. A similar behavior was found for certain type of HS binary mixtures~\cite{Odriozola11}. Thus, a glassy dynamics is expected to take place for this type of monodisperse systems~\cite{Pfleiderer08a,Pfleiderer08b}. Nonetheless, case $1.5:1$-prolates shows a $Q_6(\varphi)$ increase at $\varphi \approx 0.64$ with no sign of long range orientational order (from $p(r)$ functions), whereas case $2.0:1$-prolates shows the development of long range orientational correlations ($p(r)$ values well above zero) without changes of $Q_6(\varphi)$. The locations of these two possible (probably higher order) transitions are included in table~\ref{Table1} between brackets.     

\begin{table}
\caption{Volume density and dimensionless pressure for the different transitions as a function of the aspect ratio. $\varphi_{in}$ is the isotropic-nematic density (taken at the relative $\chi(\varphi)$ maximum) and $Z_{in}$ is its corresponding $Z$ value. $\varphi_{f}$ and $\varphi_{s}$ are the maximum and minimum densities of the fluid and solid branches, respectively. $Z_{fs}$ is the fluid-solid coexistence dimensionless pressure. A dash means that the transition was not found whereas a blank that the experiment was not carried out. Values between brackets indicate that only some measurables suggest a transition (see text). Errors are always below $3\%$ for all quantities.} \label{Table1}
\begin{tabular}{||c|c|c|c|c|c||}
\hline
\hline
aspect ratio & $\varphi_{in}$ & $Z_{in}$ & $\varphi_{f}$ & $\varphi_{s}$ & $Z_{fs}$ \\
\hline
5.0:1-prolate & 0.384 & 10.6 &  &  &  \\
4.0:1-prolate & 0.452 & 14.5 &  &  &  \\
3.0:1-prolate & 0.541 & 24.2 & 0.605 & 0.623 & 34.6 \\
2.0:1-prolate & (0.655) & (56.8) & - & - & - \\
1.5:1-prolate &  -    & - & (0.635) & (0.644) & (40.9) \\
1.3:1-prolate & 0.678 & 61.3 & 0.548 & 0.561 & 17.9 \\
1.1:1-prolate & -     & - & 0.508 & 0.538 & 13.2 \\
1:1 - sphere  & -     & - & 0.490 & 0.531 & 11.0 \\
1:1.1-oblate  & -     & - & 0.507 & 0.537 & 13.0 \\
1:1.3-oblate  & 0.682 & 69.9 & 0.543 & 0.555 & 17.1 \\
1:1.5-oblate  & -     & - & 0.662 & 0.667 & 50.3 \\
1:2.0-oblate  & 0.634 & 44.1 & - & - & - \\
1:3.0-oblate  & 0.496 & 17.5 & 0.599 & 0.618 & 29.9 \\
1:4.0-oblate  & 0.405 & 10.9 & 0.592 & 0.612 & 28.2 \\
1:5.0-oblate  & 0.344 & 8.43 & 0.582 & 0.599 & 26.2 \\
\hline
\hline
\end{tabular}
\end{table}

Table~\ref{Table1} summarizes the locations of the different transitions found as a function of the aspect ratio. This table is built by employing the histogram re-weighting technique described in refs.~\cite{Ferrenberg88,Ferrenberg89}. As pointed out by Yan and de Pablo~\cite{Yan99}, this technique is a natural complement of the REMC method, since it allows maximizing the amount of extracted information from the results of a set of simulations, i. e., form the REMC output. In our case, the histogram re-weighting scheme is given by   
\begin{equation}\label{re-w1}
h_{\beta P}(V)=\frac{h_{\beta P_0}(V) \exp(-V \Delta \beta P)}{\sum_V h_{\beta P_0}(V) \exp(-V \Delta \beta P)},
\end{equation} 
where $h_{\beta P_0}(V)$ is a reference histogram (normalized frequency against system volume) built at $\beta P_0$, $h_{\beta P}(V)$ is the re-weighted histogram evaluated at a $\beta P$ close to $\beta P_0$, and $\Delta \beta P$$=$$\beta P - \beta P_0$. Once the re-weighted histogram is obtained, the ensemble average of any function $A(V)$ at $\beta P$ can be accessed by 
\begin{equation}\label{re-w2}
<A>=\sum_V A(V) h_{\beta P}(V),
\end{equation} 
as usual. We are employing expression~\ref{re-w1} in the immediate vicinity of the discontinuous transitions (fluid-solid) to produce histograms with double peaks of equal area. From them, precise values of $\varphi_{f}$, $\varphi_{s}$, and $Z_{fs}$ are yield. For continuous transitions (isotropic-nematic) we are reporting the points which produce a local maximum in compressibility. To this end, both, expressions~\ref{re-w1} and~\ref{re-w2} with $A=\chi(V)$ are employed.      
 
\begin{figure*}
\resizebox{0.75\textwidth}{!}{\includegraphics{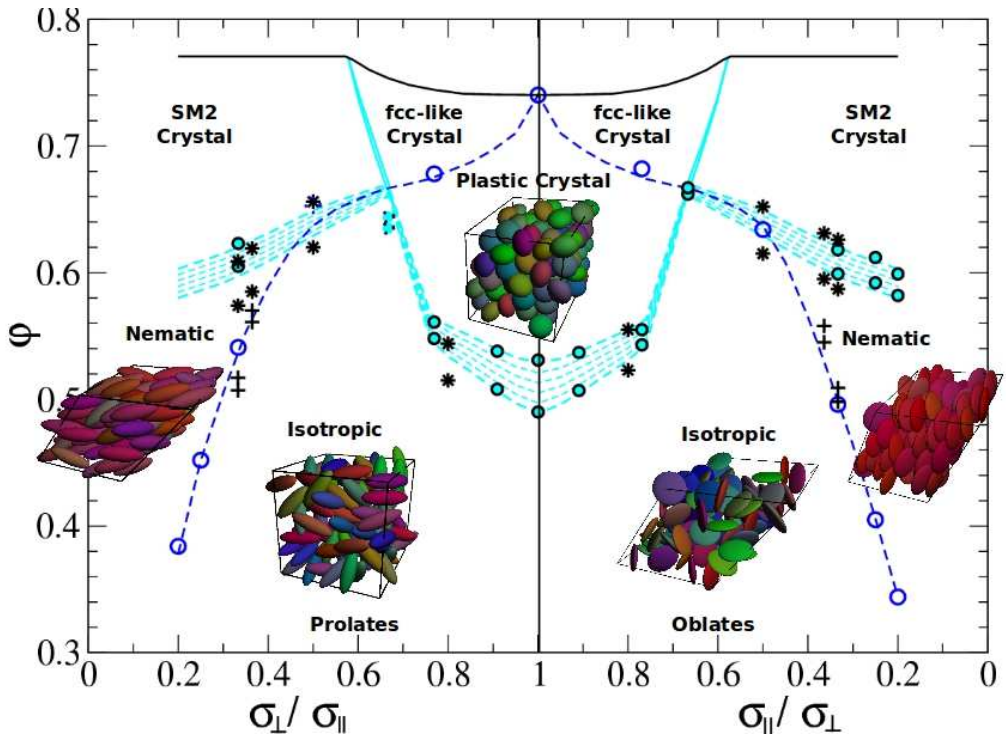}} \caption{\label{Fig5} Phase diagram of uniaxial ellipsoids. Solid circles correspond to fluid-solid transitions, whereas open circles to isotropic-nematic transitions. Symbols with dashed borders correspond to the values between brackets of table~\ref{Table1}. The black (dark) solid line is the maximally achievable density~\cite{Donev04a}. Cyan (light gray) dashed lines and blue (dark) dashed lines are guides to the eye which join the fluid-solid and isotropic-nematic transitions, respectively. The cyan (light gray) solid lines indicate fcc-SM2 transitions~\cite{Radu09}. Black (dark) plus symbols (isotropic-nematic) and asterisks (nematic-solid) are taken from the FMD~\cite{Frenkel85}. The inserted snapshots are placed according to the phase diagram region. Particles are colored according to their orientations by setting red, blue, and green to given orthogonal directions, and using a linear combination for intermediate cases. } 
\end{figure*}

The data given in table~\ref{Table1} are used to build the volume-density : aspect-ratio chart given in figure~\ref{Fig5}. This chart is divided in half by the 1:1-HS case ($\sigma_{\|}=\sigma_{\bot}$). We are placing the prolate cases at the left and oblate systems at the right of this vertical line. The particles' asymmetry increases by moving away from this central line. The extreme cases are infinitely narrow needles ($\sigma_{\bot}/\sigma_{\|} \rightarrow 0$ at the left) and infinitely thin plates ($\sigma_{\|}/\sigma_{\bot} \rightarrow 0$ at the right). At high densities, we are placing the currently accepted, maximally achievable density~\cite{Donev04a} as a black solid line. Proving this line to be the true maximally achievable density looks like a very difficult task, to say the least, taking into account that a formal proof to the Kepler conjecture (the HS maximally achievable density in three dimensional Euclidean space, i.~e., the single point at $\sigma_{\bot}=\sigma_{\|}$ and $\varphi= \pi/ \surd 18$) is still under development (the current proof is by exhaustion)~\cite{HalesFerguson}. Solid circles represent the fluid-solid transitions, whereas open circles the isotropic-nematic transitions (we used the same symbol for the plastic solid-fcc transition). The $1.5:1$ and $2.0:1$-prolate cases are shown with dashed symbols, since no formal transitions were captured but some kind of order was observed. We are also including data from the FMD~\cite{Frenkel85}. These are plus and asterisks symbols, which represent the isotropic-nematic and fluid-solid transitions, respectively. In general, there is a relatively good agreement between these data and ours. We included cyan dashed lines to join the fluid-solid transitions and blue dashed lines to join the isotropic-nematic transitions (including the plastic solid-fcc transitions). These lines should go to zero for infinitively large needles and infinitely thin plates~\cite{Frenkel85}. Furthermore, cyan solid lines were included to point out the SM2-fcc (solid-solid) transitions~\cite{Radu09}. The line at the oblate region is drawn joining the fluid-solid transition found for the $1:1.5$-oblate case and the maximally achievable density line at $\sigma_{\|}/\sigma_{\bot}=1/\surd3$. The prolate line is drawn keeping symmetry with this line around the $1:1$-HS case. These lines can be placed anywhere in the range $1.5<x<2.0$~\cite{Radu09}. All these drawn lines arise as a compromise between matching the data and producing a reasonable bordering of the phases. 

Although the chart symmetry is obvious, we should note that, in general, oblates transitions occur at lower densities than the corresponding prolate transitions. In other words, oblates are easier to rearrange in layers and crystals than prolates. Indeed, in the range $1.3<x<3.0$, where oblates yield transitions at large densities, we did not capture formal prolate transitions. These absent transitions mean that free energy (entropy) associated to fluid-like structures is similar to the corresponding free energy of the ordered state (the accessible volume gain associated to the ordering process turns relatively small in these cases). The occurrence of disordered structures at very large densities was also found experimentally~\cite{Man05} and by means of computations~\cite{Donev07,TorquatoRev}. These references report that these disordered states yield a maximally random jammed density, $\varphi \approx0.712$, peaking at $x=1.5$. They also report that prolates produce slightly higher maximally random jammed densities than oblates. This finding suggests that, at not so high densities, random states compete better with ordered states for prolates than for oblates.

\begin{figure}
\resizebox{0.48\textwidth}{!}{\includegraphics{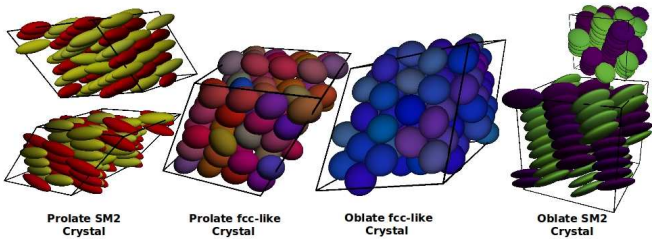}} \caption{\label{Fig6} Snapshots showing the obtained prolate SM2 crystal, prolate fcc, oblate fcc, and oblate SM2, as labeled. The two leftmost and the two rightmost panels show different views of the same snapshots. Particles are colored according to their orientations. For SM2 structures colors correspond to a binary population of orientations (see text). For the fcc structures, red, blue, and green colors are set to given orthogonal directions and a linear combination is used for intermediate cases.} 
\end{figure}

Snapshots for the crystal-like structures developed at high pressures are shown in figure~\ref{Fig6}. These are SM2 structures, for $3:1$-prolates and $1:4$-oblates, and fcc-like structures. The particles belonging to the fcc-like structures are colored by setting red, blue, and green to given arbitrary orthogonal directions, and using a linear combination for intermediate cases. Thus, similar colors mean that particles' main axes point to similar directions. The same procedure was applied for the embedded snapshots of figure~\ref{Fig5}. Consequently, many different colors appear in isotropic phases, whereas similar tones arise in nematic phases. In particular, the plastic solid structure shown in figure~\ref{Fig5} (for $1:1.3$-oblates at not very high pressures) shows different colors, pointing out the absence of long range orientational order. However, the fcc-like structures in figure~\ref{Fig6}, for $1.3:1$-prolates and $1:1.3$-oblates at pressures above the plastic solid-fcc transitions, show homogeneous reddish and bluish tones, respectively. Here it is also seen that, at similar pressures, oblates align better than prolates (the bluish tones look more homogeneous than the reddish ones). Finally, we should say that the particles' orientations of these fcc-like structures are not expected to perfectly align at higher pressures since this would impede the systems reaching densities above $\varphi= \pi/ \surd 18$~\cite{TorquatoRev}.     

As explained in ref.~\cite{Pfleiderer07}, the SM2 family of structures consists of the replication of a two-particle monoclinic cell, while keeping a nonzero angle between the principal axes of these particles. The replication of this simple cell produces layered patterns, where the intra-layer structure is constituted by columns of aligned particles. For the snapshots at the left and right of figure~\ref{Fig6} ($3:1$-prolates and $1:4$-oblates, respectively) we observed the principal axes of the particles produce a clear double-peak distribution of their projections on a plane perpendicular to the stacks (columns) average direction. By painting both populations with different colors the layered pattern clearly appears, as can be seen for both, prolates and oblates in figure~\ref{Fig6}. The average angle between the two populations is close to 9 and 13 degrees for $3:1$-prolates and $1:4$-oblates, respectively. We should also mention that, by applying a similar analysis to the structures reported in previous works~\cite{OdriozolaHE,GuevaraHE}, we also obtained the double peak distribution. In these cases the average angles differ $7-11$ degrees from one another. Hence, both tilted columnar phases described in previous works~\cite{OdriozolaHE,GuevaraHE} also belong to the SM2 family. 

As pointed out, some particular structures of this family were shown to produce the highest, to the actual knowledge, achievable density~\cite{Donev04a}. Nonetheless, there is no conclusive evidence that $\varphi=0.7707...$ is the highest achievable density or that the SM2 family should produce it~\cite{Donev04a}. Thus, the fact that the REMC algorithm is naturally producing the SM2 crystal for both, prolates and oblates with a large enough asymmetry, supports the idea that the SM2 arrangement is, in fact, the structure that yields the highest achievable density~\cite{Donev04a} (otherwise at least another solid-solid phase transition should occur at very high densities).

\section{Conclusions}
The existence of a recently found family of structures (SM2)~\cite{Pfleiderer07} which yields the highest maximum achievable density~\cite{Donev04a} and a free energy below that of a stretched-fcc for sufficiently anisotropic ellipsoids~\cite{Radu09} motivated us to revisit the phase diagram of uniaxial ellipsoids. For this purpose, we implemented replica exchange Monte Carlo simulations, which are known to improve sampling at high pressure conditions. Thus, solids phases are accessed from loose and random configurations. 

The method produced a certain type of SM2 structure for $x\!:\!1$-prolates and $1\!:\!x$-oblates with $x\geq3$, strongly suggesting they are preferred at equilibrium. This result also supports the idea that SM2 arrangements lead to the highest achievable density~\cite{Donev04a}, since, otherwise, a solid-solid transition should take place at even larger densities (in our view, this would be odd but possible). It also produced fcc-like structures showing long range orientational order at very high pressures. This, although expected~\cite{Frenkel85}, is hard to capture. We also found that oblates and prolates with $x=1.5$ and 2 hinder the formation of ordered structures. Finally, we should mention that the locations of the phase transitions relatively agree with those given in the Frenkel-Mulder diagram.

\begin{acknowledgments}
The author thanks project No.~Y.00119 SENER-CONACyT for financial support and fruitful discussions with Drs.~Gurin Peter and Arturo Moncho-Jord\'a.
\end{acknowledgments}


%

\end{document}